\documentclass{article}
\usepackage{emulateapj,epsfig}

\slugcomment{Accepted for publication in ApJ Lett}
\makeatletter

\newenvironment{inlinefigure}{%
\def\@captype{figure}%
\noindent\begin{minipage}{0.999\linewidth}\begin{center}}
{\end{center}\end{minipage}\smallskip}
\makeatother

\begin{document}

\title{Early Structure Formation and Reionization\\ 
in a Warm Dark Matter Cosmology}

\author{Naoki Yoshida, Aaron Sokasian, Lars Hernquist}
\affil{Harvard-Smithsonian Center for Astrophysics, 60 Garden Street,
Cambridge, MA 02138}
\author{Volker Springel}
\affil{Max-Planck-Institut f\"ur Astrophysik, Karl-Schwarzschild-str. 1,
Garching bei M\"unchen}

\begin{abstract}

We study first structure formation in $\Lambda$-dominated universes
using large cosmological $N$-body/SPH simulations. We consider a
standard $\Lambda$CDM model and a $\Lambda$WDM model in which the mass
of the dark matter particles is taken to be $m_{X}=10$ keV.  The
linear power spectrum for the $\Lambda$WDM model has a characteristic
cut-off at a wavenumber $k=200$ Mpc$^{-1}$, suppressing the formation
of low mass ($< 10^6 M_{\odot}$) nonlinear objects early on.  The
absence of low mass halos in the WDM model makes the formation of
primordial gas clouds with molecular hydrogen very inefficient at high
redshifts.  The first star-forming gas clouds form at $z\approx 21$ in
the WDM model, considerably later than in the CDM counterpart, and the
abundance of these gas clouds differs by an order of magnitude between
the two models.  We carry out radiative transfer calculations by
embedding massive Population III stars in the gas clouds.  We show
that the volume fraction of ionized gas rises up close to 100\% by
$z=18$ in the CDM case, whereas that of the WDM model remains
extremely small at a level of a few percent. Thus the WDM model with
$m_{X}=10$ keV is strongly inconsistent with the observed high optical
depth by the {\it WMAP} satellite.

\end{abstract}


\section{Introduction}

Popular cosmological models based on Cold Dark Matter (CDM) predict
that the first stars form in low mass ($\sim 10^6 M_{\odot}$) dark
halos at redshifts $z\approx 20-30$ when primordial gas condenses via
cooling by hydrogen molecules (Abel, Bryan \& Norman 2002; Bromm,
Coppi \& Larson 2002).  Hierarchical structure formation eventually
leads to the emergence of a population of early generation stars
(Yoshida et al. 2003), which may have at least partly reionized the
Universe soon after the end of the Dark Ages.  The first year {\it
WMAP} result of the measurement of CMB polarization implies a large
optical depth $\tau_{e}=0.17\pm 0.04$, indicating that reionization
could have occurred as early as $z_{\rm reion}\sim 17$ (Kogut et
al. 2003).  The theoretically predicted formation epoch of the first
stars in CDM models thus appears plausible in light of the {\it WMAP}
result, and indicates that an early generation of stars may have
contributed to reionization.  On the other hand, from the theoretical
point of view, it is intriguing and important to ask whether or not
such an early reionization epoch is compatible in detail with models
other than CDM, or even with CDM itself.

Warm Dark Matter (WDM) models predict exponential damping of the
linear matter power spectrum on small length scales (Bardeen et al. 1986).
The characteristic length scale is given by the free-streaming length
of the warm dark matter particle as, $R_{\rm
f}=0.31(\Omega_{X}/0.3)^{1/3}(h/0.65)^{1.3} (m_{X}/{\rm keV})^{-1.15}$
$h^{-1}$Mpc, where $m_{X}$ is the particle mass (Bode, Ostriker \&
Turok 2001).  Owing to the suppression of power on small scales, the
abundance of low mass halos in WDM models is considerably smaller than
in CDM models, perhaps resulting in better agreement with the observed
matter distribution on sub-galactic scales (Bode et al. 2001).  
Various constraints have been placed on the mass $m_{X}$.
Based on an analysis of the clustering properties of the Ly-$\alpha$
forest, Narayanan et al. (2000) conclude that the lower limit on the
mass is $m_{X}=0.75$ keV.  Dalal \& Kochanek (2002) used lensing
statistics and found that models with $m_{X} < 5$ keV are incompatible
with the observed abundance of substructure in distant
galaxies. Barkana, Haiman \& Ostriker (2001) concluded that models
with $m_{X} < 1$ keV are likely to be ruled out {\it assuming} the
reionization redshift $z_{\rm reion}\sim 6$. Reionization at an
earlier epoch as implied by the {\it WMAP} data generically requires
early structure formation, and thus may place a more stringent
constraint on the mass of dark matter $m_{X} \gg 1$ keV (Somerville,
Bullock \& Livio 2003).

In this {\it Letter}, we explore the formation of the first baryonic
objects in a cosmological model in which dark matter is warm, rather
than cold.  Since models with $m_{X}\lesssim 1$ keV appear to be
inconsistent with an array of observations, we consider a model with
$m_{X}=10$ keV.  Although the motivation from particle physics for
elementary particles with such an intermediate mass is somewhat
unclear (but see Kawasaki, Sugiyama \& Yanagida 1997; Bode et al. 2001), 
it is important to examine the effect of suppressing
the linear power spectrum on early structure formation.  We
specifically study the formation of primordial gas clouds and of their
host halos with dark masses $10^5-10^6 M_{\odot}$ in a $\Lambda$CDM
and in a $\Lambda$WDM universe, and compare the results for the two
models. While the currently available observations do not directly probe
structure in the redshift range we consider here, future CMB
polarization experiments will ultimately reveal how and when the
Universe was reionized (Kaplinghat et al. 2003), and thus will be able
to distinguish the structure formation histories predicted by the two
models.

\begin{inlinefigure}
\resizebox{8cm}{!}{\includegraphics{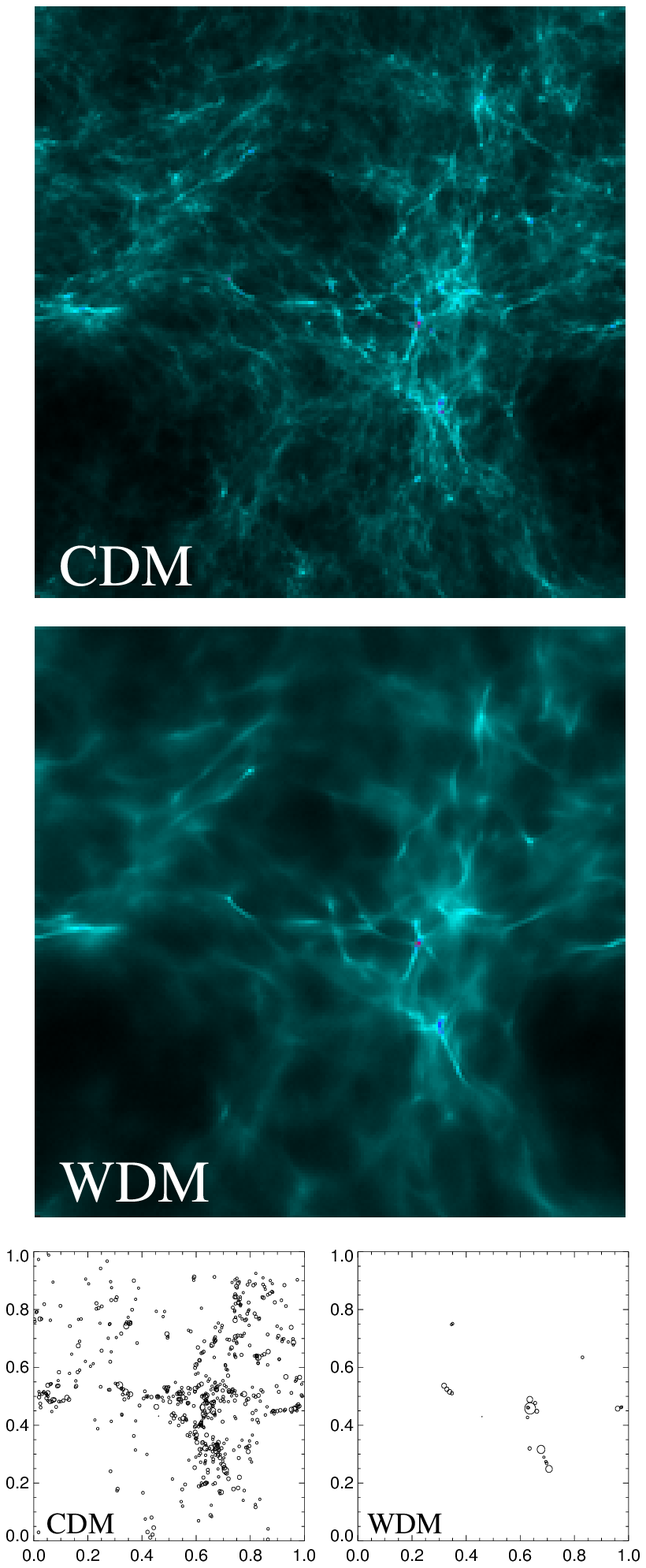}}
\caption{The projected gas distribution in the CDM (top)
and the WDM (middle) simulations at $z=20$. Figures in the bottom portion show
the distribution of dark halos with mass greater than $10^5 M_{\odot}$
for the CDM (left) and for the WDM (right) model. \label{plot1}}
\end{inlinefigure}

\section{The $N$-body/SPH simulations}

We use the parallel Tree-PM/SPH solver GADGET2 in its fully
conservative entropy form (Springel \& Hernquist 2002).  We follow the
non-equilibrium reactions of nine chemical species (e$^{-}$, H, H$^+$,
He, He$^{+}$, He$^{++}$, H$_{2}$, H$_{2}^{+}$, H$^{-}$) using the
reaction coefficients compiled by Abel et al. (1997). We use the
cooling rate of Galli \& Palla (1998) for molecular hydrogen.  We
study both CDM and WDM models with matter density $\Omega_{0}=0.3$,
baryon density $\Omega_{\rm b}=0.04$,
cosmological constant $\Omega_{\Lambda}=0.7$ and expansion rate at the
present time $H_{0}=70$km s$^{-1}$Mpc$^{-1}$. We set the index of the
primordial power spectrum $n_{s}=1$ 
\begin{inlinefigure}
\vspace*{5mm}
\resizebox{9cm}{!}{\includegraphics{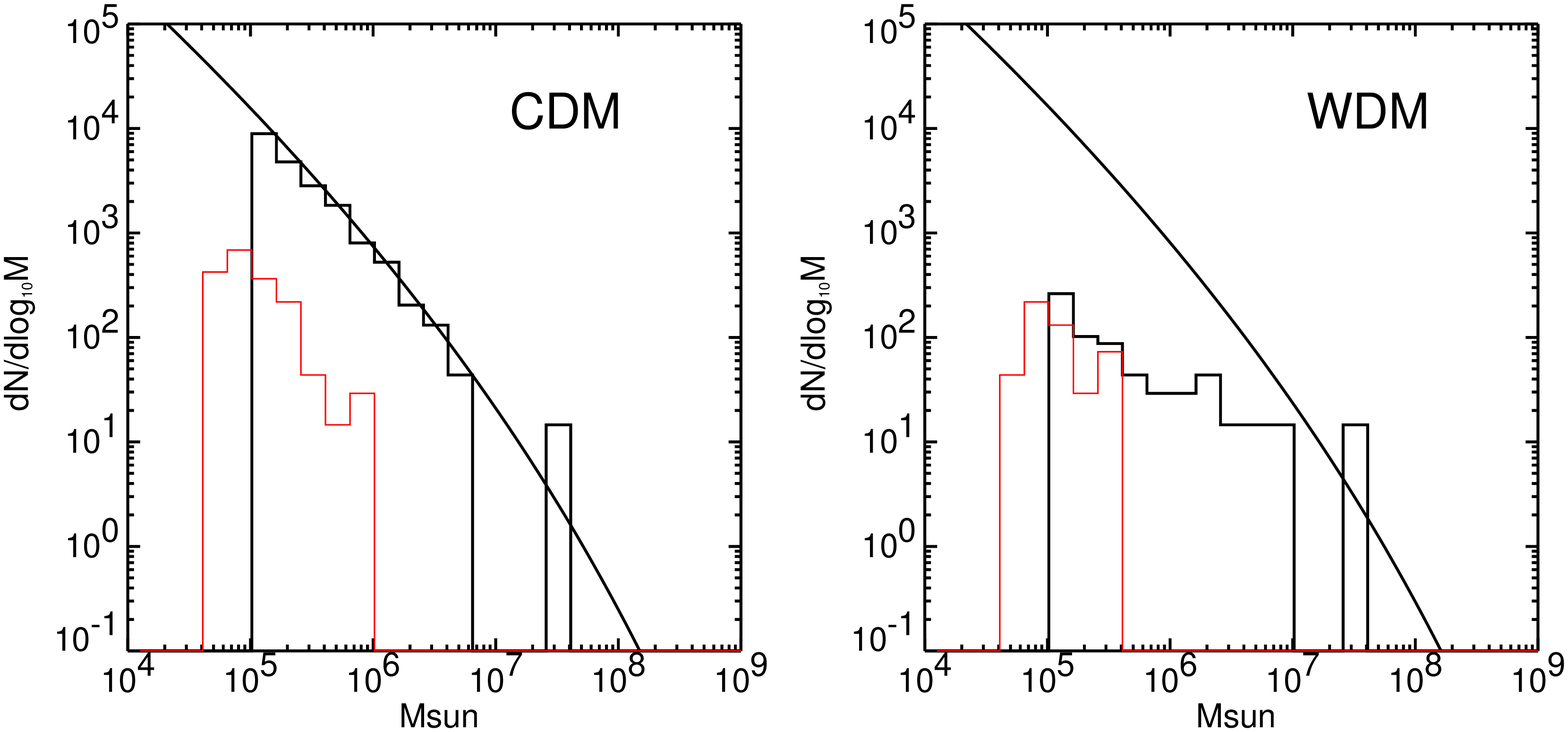}}
\caption{Mass function of dark matter halos at $z=20$.
The solid line is the Press-Schechter mass function computed 
for the CDM model, which agrees remarkably well with our CDM simulation.
The thin histograms show the abundance of all the subhalos found in the 
simulation box. 
\label{plot2}}
\end{inlinefigure}
\\
and normalize the fluctuation
amplitude by setting $\sigma_8=0.9$.  
%We follow Bode et al. (2001) to compute the transfer function
%for the WDM model, assuming the dark matter mass $m_{X}$=10 keV.  
%
% revise
We follow Bode et al. (2001, see their Appendix A) to set-up the initial condition
for the WDM model, assuming the dark matter mass $m_{X}$=10 keV.  
In order to avoid spurious clumping in the initial particle set-up (see
G\"otz \& Sommer-Larsen 2002), we use ``glass'' particle
distributions.  Further simulation details are given in Yoshida et
al. (2003, hereafter Paper I).  Both of the simulations employ
$2\times 324^3$ particles in a cosmological volume of 1 Mpc on a
side. The mass per gas particle is then 160 $M_{\odot}$, whereas the
mass of the dark matter simulation particles is 1040 $M_{\odot}$.  In
Paper I, we carried out numerical convergence tests using higher
resolution simulations and concluded that the mass resolution adopted
here is sufficient to follow the cooling and collapse of primordial
gas within low mass ($\sim 10^6 M_{\odot}$) halos.

\section{Results}

Figure 1 shows the projected gas distribution at $z=20$ for the two
models.  Panels in the bottom portion show the distribution of dark
matter halos in each simulation.  The effect of the exponential
cut-off in the initial power spectrum is evident in Figure 1; the gas
distribution in the WDM model is much smoother than in the CDM case.
Also, the abundance of low mass halos crucial for the formation of
primordial gas clouds is significantly reduced in the WDM model.  We
locate the dark matter halos by first running a friends of friends
(FOF) groupfinder with linking parameter
$b=0.164$.  We discard halos which consist of fewer than 100 particles.
We then use the SUBFIND algorithm (Springel et al. 2001) which
identifies gravitationally self-bound sets of particles that are at a
higher density than the smooth background.  We carry out the
latter step, because, particularly in WDM models, filamentary
structures tend to be identified as halos and such objects often
contain many gravitationally bound ``subhalos'' (Knebe et al. 2002).
Using this two-step method, we can robustly identify gravitationally
bound objects in our simulations.  We compare the mass function of the
dark halos in Figure 2.  There we also show the
abundance of {\it all} the subhalos identified in the simulation box
by thin histograms.  The mass function for the CDM simulation is
well-fitted by the Press-Schechter mass function (solid line), 
as found by Jang-Condell \& Hernquist (2001).  
We compare this analytic mass function for the CDM model 
\begin{inlinefigure}
\resizebox{8.5cm}{!}{\includegraphics{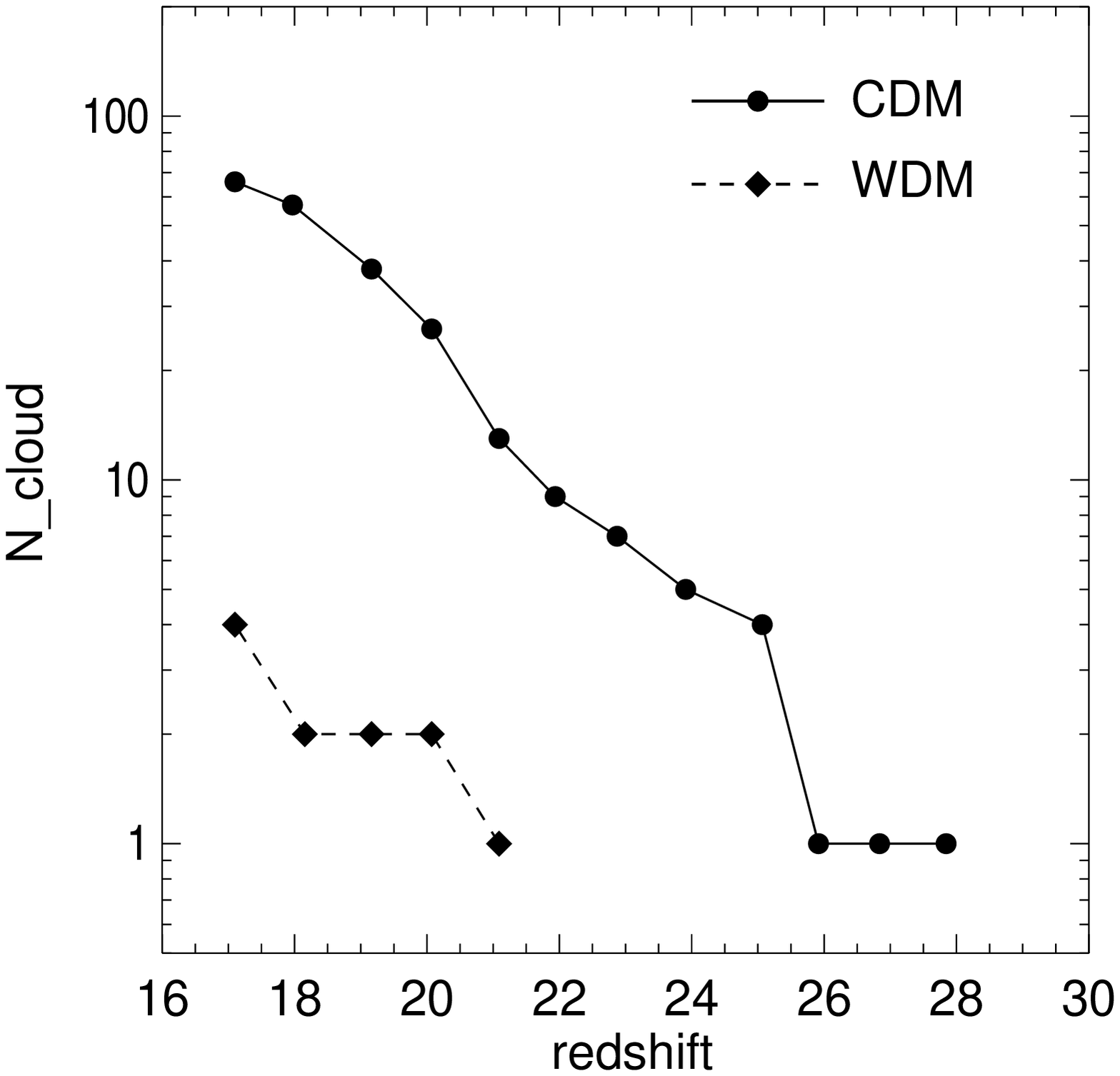}}
\caption{The number of star-forming gas clouds
as a function of redshift. \label{plot3}}
\end{inlinefigure}
\\
with the result of the WDM
simulation (right panel). The difference is nearly two orders of
magnitude at a mass scale of $10^5 M_{\odot}$. Note, however, that the difference
in the mass function between the two models becomes {\it smaller} on
larger mass scales, confirming that the suppression of the linear
power spectrum in the WDM model affects only small mass scales.  
%We also note that the most massive halos in both models have a
%virial temperature higher than 8000 K at $z=20$. 
%Since we do not
%include photo-dissociating radiation, the gas in these halos has been
%cooled predominantly via molecular hydrogen cooling, rather than via
%atomic hydrogen transitions.  
%For simplicity, we include the gas
%clouds within these large halos as sites for Population III star
%formation.

In Figure 3, we plot the number of gas clouds against redshift.  We
define groups of cold ($T<500$K), dense ($n_{\rm H} > 500$cm$^{-3}$)
gas particles as ``gas clouds''.  In order to locate dense gas clumps,
we again run a FOF groupfinder with
$b=0.05$ to gas particles. From each group we discard gas particles which do not
satisfy the above criteria. We then identify the group as a
star-forming gas cloud if the cold gas mass exceeds $M_{\rm
Jeans}=3000 M_{\odot}$.  The first gas cloud is identified in this
manner at $z=28$ in the CDM model, whereas it is much later at $z=21$
in the WDM model. The total number of gas clouds in the simulated
volume differs by about an order of magnitude
in the redshift range plotted.  At $z=20$, we identified 26 gas clouds
in the CDM model, and there are only 2 gas clouds found in the WDM
case.  The corresponding numbers at $z=17$ are 66 and 4 for the CDM
and WDM models, respectively.  Note that the number of gas clouds
does not represent the {\it true} abundance of the first stars,
because our simulations do not include all feedback processes from
the first stars.

An important question is how reionization by an early
generation of
stars proceeds at high redshift in the two
models.  To address this
question, we carry out radiative transfer simulations for a specific
model of star formation, as follows.  The first stars formed out of a
chemically pristine gas are likely to be very massive (Abel et
al. 2002; Omukai \& Palla 2003).  As in Paper I, we assume that each
gas cloud forms a single massive star, following the usual ``one star
per halo'' assumption for ``mini-halos.''  As our fiducial model, we
set the mass of a Population III star to be 300 $M_{\odot}$ with a
lifetime of 3 million years, and also assume a constant photon escape
fraction $f_{\rm esc}=1$. 
%Although the high escape fraction might seem
%unrealistic, it is indeed {\it plausible} for massive population III
%stars in mini-halos (Tom Abel, private communication).  
%
% revised
Although the high escape fraction might seem
unrealistic, it may indeed be plausible because 
the gas in mini-halos 
\begin{inlinefigure}
\resizebox{8.5cm}{!}{\includegraphics{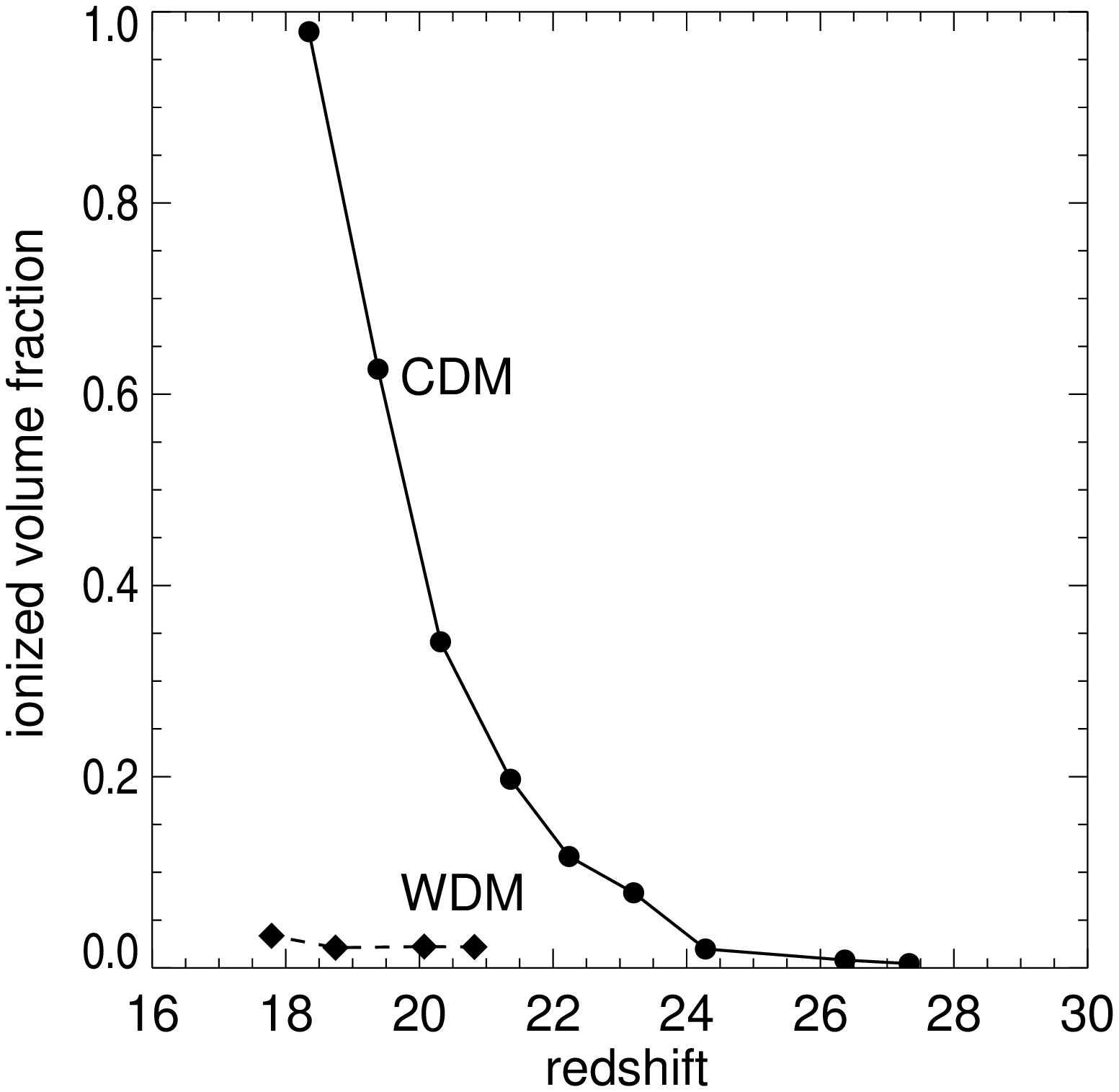}}
\caption{The volume fraction of ionized regions.\label{plot4}}
\end{inlinefigure}
\\
can be wholly
ionized and photo-evapolated by a single massive star 
assuming the gas distribution is reasonably smooth (Oh et al. 2001).  
%revise end
We then run
multi-source radiative transfer simulations using the technique of
Sokasian et al. (2003). 
%Briefly, the code utilizes a gas density field
%
%revise
Briefly, the code utilizes a post-procesed gas density field
%revise end
defined on a $200^3$ grid by casting multiple rays from sources in
an adaptive fashion. Photon absorption and recombination are computed
along the rays, and each cell carries its own properties such as
clumping factor and ionization fraction. The gas evolution between two
adjacent outputs, most importantly recombination, is computed on a
cell-by-cell basis using these quantities. 
% Details of the ray-tracing
% scheme can be found in Sokasian et al. (2001, 2003).
%
% revise
Note that our ray-tracing simulations employ a {\it one-step}
scheme (Sokasian et al. 2001) in which the gas density evolution
due to radiative feedback is not taken into account. 
In order to mimic strong radiative feedback within HII regions,
we implemented a ``volume exclusion effect'' by disabling sources if they
lie within already ionized regions.  
In practice, we turn on a source
only if the ionization fraction of its surrounding gas is below 0.05.
We compute the total volume filling factor of the ionized medium
from the output of the ray-tracing simulation.  Figure 4 shows
the evolution of the volume filling factor computed in this manner for
the two models. Initially, there is only a single HII region and so the
filling factor is very small. As more stars are formed, the filling
factor rapidly increases close to 1.0 by $z=18$ in the CDM model,
causing complete reionization. On the other hand, due to the
small number of sources, the ionized volume fraction in the WDM model
remains extremely small, only up to 0.03 at $z=18$.
 
\section{Summary and Discussion}

The suppression of small scale power in the WDM model has a
significant impact on the formation of primordial gas clouds.
Hierarchical growth of halos with mass $10^5-10^6 M_{\odot}$ is not
seen in the WDM model, and gas cloud formation is nearly completely
suppressed until halos with mass $\sim 10^7 M_{\odot}$ collapse at
$z\sim 20$. The global star-formation rate thus remains very
small at $z>17$ (see Figure 3), regardless of the details of star formation.  Our
radiative transfer calculations show that reionization by early
Population III stars is a very slow and inefficient process in the WDM
model.  To be compatible with the observed high optical depth of WMAP,
the ionization fraction in the WDM model {\it must} increase rapidly at
$z<18$. As Sokasian et
al. (2003) argue, the optical depth can be as high as $\tau_{\rm e}
\sim 0.15$ only if a large number of ionizing photons are produced in
(proto-)galaxies at $z<18$.  Clearly the WDM model we consider here is
disfavored, if not ruled out, in light of the {\it WMAP} results.
WDM models with $m_{X}\lesssim 1$ keV are likely to be ruled out as
argued by Barkana et al. (2001) and Somerville et al (2003), whereas
models with $m_{X}=100$ keV would be essentially indistinguishable
from CDM models in the context of structure formation.

We now turn to the question of whether or not the CDM model is
compatible with the observed high optical depth.  Using the results of
our ray-tracing calculation combined with that of Sokasian et
al. (2003) for Population II sources, we compute $\tau_{e}$ as a function 
of redshift. We combine the
contributions from the two modes of star-formation in different
redshift intervals assuming that the onset of Population II begins
exclusively at $z<18.5$ {\it and} that there are no Population III 
stars since then. While this is clearly an
over-simplification, it provides a conservative estimate for the total
optical depth. To this end we re-run the $f_{\rm esc}=0.20$
model from Sokasian et al. (2003) with the initial condition that the
IGM was fully reionized by $z\simeq18.5$ and was uniformly heated 
to a temperature of $1.5\times 10^{4}$ K. Figure 5 shows $\tau_{e}$ for the
CDM model computed in this manner.
Note the slope of the curve decreases at $15 < z <18$,
reflecting the decline in the ionization fraction owing to
recombinations at a time when the emissivity from Population II
sources is still low. Population III stars in the CDM model give a $\Delta\tau\sim$0.05
whereas the ordinary stellar populations 
contribute $\Delta\tau\sim$0.09 at $z<15$, giving a total of $\tau \sim 0.14$, 
in good agreement with the {\it WMAP} result. 
In Figure 5, we also show $\tau_{e}$ for the case with only the 
contribution from Population II sources.
The reionization history for the WDM model would be close to this
case, with the contribution from Population III stars in 
%We note that, although our model on the massive Population III star formation is
%simple, it is well motivated by recent theoretical studies
%(Abel et al. 2002; Bromm et al. 2002; Omukai \& Palla 2003).  
\begin{inlinefigure}
\resizebox{8.5cm}{!}{\includegraphics{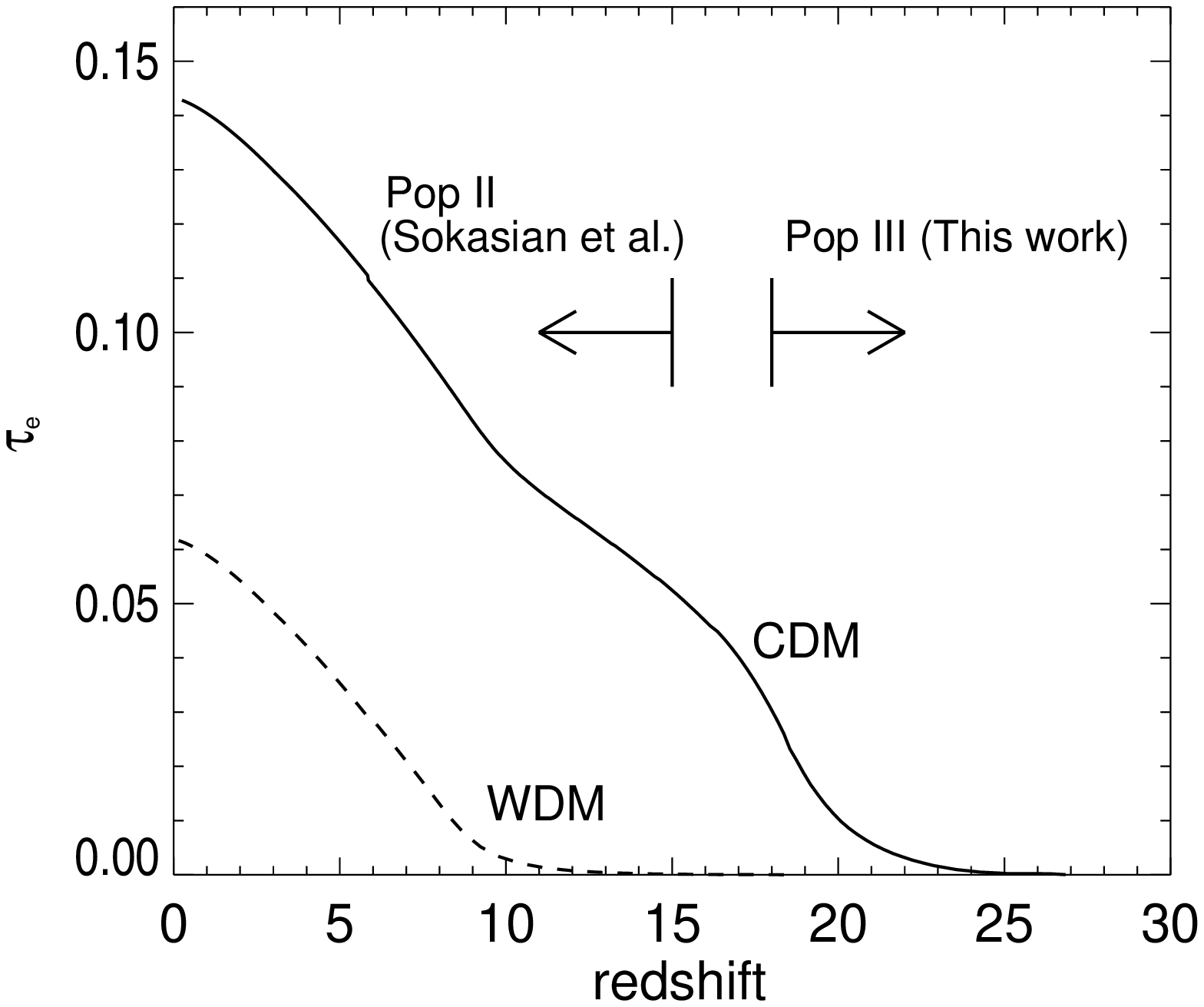}}
\caption{Thomson optical depth as a function of redshift.
We compute $\tau_{e}$ using the combined result of the present work for Pop
III ($z>18$) and
Sokasian et al. (2003) for Pop II ($z<15$) .\label{plot5}}
\end{inlinefigure}
\\
mini-halos being negligible.
In a forthcoming paper, we will study extensively a number of models 
using a more detailed prescription of early star-formation.  The
reionization history could be complex as in the double reionization
scenario (Cen 2002; Wyithe \& Loeb 2003), if a dramatic transition between the two modes of
star-formation, Pop III to Pop II, occurs at $6<z<18$.  
While it is yet too early to draw a definite conclusion, given the uncertainty
in the {\it WMAP} measurement of the optical depth,
future CMB polarization
measurements such as {\sl Planck} will probe the ionized hydrogen
fraction at high redshift as well as the total optical
depth accurately (Kaplinghat et al. 2003), and thus will be able to place a
strong constraint on the structure formation scenario and on the mass
of dark matter.
\\

NY thanks Paul Bode for helpful comments on the initial set-up 
of the WDM simulation.
This work was supported in part by NSF grants ACI 96-19019, AST 98-02568, 
AST 99-00877, and AST 00-71019.

\end{document}